\documentclass[allclo]{FBSart}
\usepackage{epsfig,psfig}
\title{The pion electromagnetic 
form-factor in a QCD-inspired model
\footnote{Presented at {\it Light-Cone 2004}, 
Amsterdam, The Netherlands, August 16-20, 2004.
}}
\author{$^a$$^{,b}$J.~P.~B.~C.~de~Melo, $^c$T.~Frederico,
$^d$E.~Pace and $^e$G.~Salm\'e }
\institute{ $^a$Instituto de F\'\i sica Te\'orica, 
Universidade Estadual Paulista,
01405-900, S\~ao Paulo, SP., Brazil 
\newline
$^b$Centro de Ci\^encias Exatas e Tecnol\'ogicas, 
Universidade Cruzeiro do Sul \\
08060-070, S\~ao Paulo, SP., Brazil
\newline
$^c$Dep. de F\'\i sica, ITA, CTA,
12.228-900, S\~ao Jos\'e dos Campos, S\~ao Paulo, Brazil
\newline
$^d$Dipartimento
di Fisica, Universit\`a di Roma "Tor Vergata",
and INFN, Sezione
Tor Vergata, Via della Ricerca Scientifica,
1-00133,Roma, Italy
\newline
$^e$INFN, Sezione Roma I, Pl.le A. Moro 2, 1-00185,
Roma, Italy}
\runningauthor{J. P. B. C. de Melo et al} \runningtitle{LC 2004}
\begin{document}
\maketitle
\vspace{-0.4cm}
\begin{abstract}
We present detailed numerical results for the pion space-like
electromagnetic form factor obtained within a recently proposed
model of the pion electromagnetic current in a confining
light-front QCD-inspired model. The model incorporates the vector
meson dominance mechanism at the quark level, where the dressed
photon with $q^+>0$ decays in an interacting quark-antiquark pair,
which absorbs the initial pion and produces the pion in the final
state.
\end{abstract}
\vspace{-0.4cm}
\section{Introduction}

The light-front dynamics~(LFD)~\cite{Dirac49} is a suitable framework for
describing electromagnetic~(em) interactions and bound states for
hadronic systems~(\cite{Brodsky98}). Within light-front dynamics
and in the framework of an inspired-QCD model~\cite{Pauli2001}, we
have studied the pion em form factors in both space- and
-time-like regimes~\cite{Pacheco2004}. In this contribution,
detailed numerical results are presented and  compared to
the experimental data
the electromagnetic form factor for the pion in the space-like
region for $-q^2~<~10~(GeV/c)^2$.

In order to calculated the matrix elements of the pion em current
we use the Mandelstam formula. In the space-like (SL) region one has
 \begin{eqnarray} j^{\mu}
~=&&-\imath e ~2 \frac{m^2}{f^2_\pi} N_c\int \frac{d^4k}{(2\pi)^4}
\overline{\Lambda}_{\pi'}(k+P_{\pi},P_{\pi^{\prime}})
\Lambda_{\pi}(k,-P_{\pi}) \nonumber \\ && \times
Tr\left[S(k+P_{\pi}) \gamma^5 S(k-q) ~\Gamma^\mu(k,q)~S(k)
\gamma^5 \right] \ , \label{amande}
\end{eqnarray} where $S(k,p)$ is Dirac propagator for the quarks,
$m$ is constituent quark mass, $N_c=3$ is the number of colors and
the factor 2 in the equation above, comes from isospin algebra.
$\Gamma^{\mu}(k,q)$ is the quark-photon vertex, $q^{\mu}$   the momentum
transfer and $\bar{\Lambda_{\pi}}$   the pion vertex function.
The expression of the current appropriate for the time-like
(TL) region can found in \cite{Pacheco2004}.

In principle, the complete vertex function for the photon and the
pion should be obtained by solving an inhomogeneous and a
homogeneous Bethe-Salpeter equation, respectively. In general, the
pion vertex can contain a pseudovector spinor operator, besides
the pseudoscalar one shown in Eq.(\ref{amande}) and adopted in what follows.

The relative light-front time in Eq.~(\ref{amande}) is eliminated
by performing the integration over $k^-~(k^0-k^3)$ in a frame
where $q^+~\neq 0$ ($q^+=q^0+q^3$) and $\vec q_{\perp}=\vec
P_{\pi^\prime\perp}=\vec P_{\pi\perp}=\vec 0_\perp$ ~\cite{Lev98}.
We disregard the analytic structure of the vertex functions when
performing the $k^-$ analytical integration. To further simplify
the model calculations, the pion is considered massless. Due to
the frame choice and the hypothesis of a massless pion,
only the pair production
mechanism survives in the photon absorption process.
Mathematically the pair diagram comes from the residue of the
integrand in Eq.~(\ref{amande}) evaluated at
$k^-=q^--(q-k)^-_{on}$ (the on-minus-shell value) for $0<k^+<q^+$.

The hadronic part of the virtual-photon wave function is dominated by
the vector-meson (VM) contribution, and this leads to a model for the
quark-photon vertex given by a sum over the VM vertex function
 properly weighted. The momentum components of
VM and pion vertex functions, in a constituent quark-antiquark
approach, come from the front-form wave functions in a QCD-inspired
model that describes the meson spectrum in both the light- and
heavy-quark sectors~\cite{Pauli2001}.

\section{Pion electromagnetic form factor}

The electromagnetic form factor of the pion is the invariant
defined from the general form of the matrix elements of the em
current, as:
\begin{eqnarray}
j_{SL}^{\mu}  & = & \langle \pi | \overline{q} \gamma^{\mu} q |
\pi^{\prime}  \rangle = (P^{\prime\mu}_\pi+P^{\mu}_\pi)
F_{\pi}(q^2) \ ,
\nonumber \\
j_{TL}^{\mu} & = & \langle \pi \overline{\pi} | \overline{q}
\gamma^{\mu} q| 0 \rangle = (P^{\prime\mu}_\pi-P^{\mu}_\pi)
F_{\pi}(q^2) \ , \label{ffactors}
\end{eqnarray}
in the space-like and time-like regions, respectively.

The integration of the Mandelstam formula Eq.~(\ref{amande})
requires the pion and the quark-photon vertex functions, which are
evaluated within LFD. In the following, we present the main
ingredients of our model calculation and refer to
\cite{Pacheco2004} for further details.

The quark photon vertex in Eq.~(\ref{amande}) is constructed using
the vector meson dominance (VMD) hypothesis. In the case of the
plus component of the em current the quark-photon vertex is
\begin{eqnarray}
\Gamma^+(k,q) = \sqrt{2} \sum_{n, \lambda} \left [
\epsilon_{\lambda} \cdot \widehat{V}_{n}(k,k-q)  \right ]
\Lambda_{n}(k,P_n) ~ { [\epsilon ^{+}_{\lambda}]^* f_{Vn} \over
\left [ q^2 - M^2_n + \imath M_n \Gamma_n(q^2)\right ]} \ ,
\label{cur5}
\end{eqnarray}
 where $f_{Vn}$ is the decay constant
of the n-th vector meson with mass $M_n$ and quadrimomentum
$P_{n}$. $\Gamma_{n}(M^2_n)$ is the total width and
$\epsilon_{\lambda}$ is the polarization of the vector meson. In Eq.
(\ref{cur5}),
  $\epsilon_{\lambda} \cdot
\widehat{V}_{n}(k,k-q) \Lambda_{n}(k,P_n)$ is the VM vertex
function, including the operator structure in the spinor space. We
use the operator form $\widehat{V}^{\mu}_{n}(k,k-q) = \gamma^{\mu}
- (k^{\mu}_{on}-(q-k)_{on}^{\mu} )/( M_{0} + 2 m )$
~\cite{Jaus90}, where for simplification all momentum are
evaluated on-minus-shell and $M_0$ is the free mass of the virtual
$q\bar{q}$ pair produced by the decay of the VM resonance.

Introducing Eq.~(\ref{cur5}) in Eq.~(\ref{amande}), the
electromagnetic form factor is written as
\begin{eqnarray}
F_{\pi}(q^2) = \sum_n~ {f_{Vn} \over q^2 - M^2_n + \imath M_n
\Gamma_n(q^2)} ~
 g^+_{Vn}(q^2)\ ,
\label{tlff}
\end{eqnarray}
 where $g^+_{Vn}(q^2)$, for $q^2 > 0$,
is the form factor for the VM decay in a pair of pions, and for
$q^2<0$ is the form factor for the photon absorption through the
coupling with a VM resonance.

\begin{figure}[t]
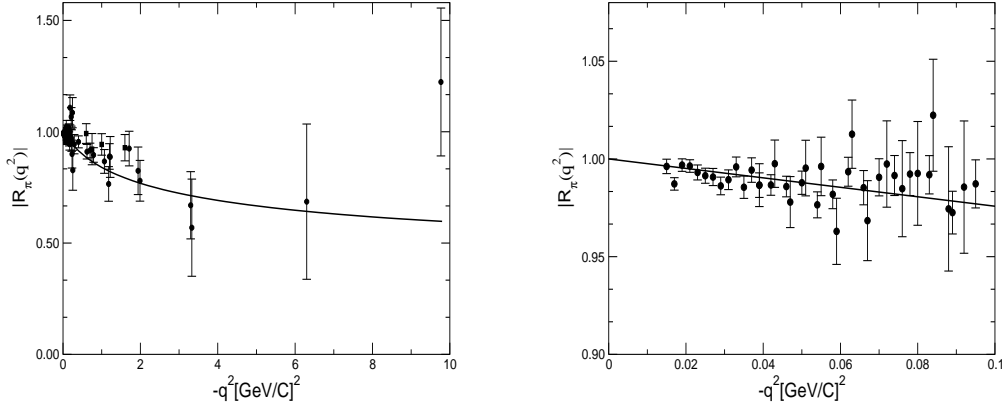

\begin{center}
\psfig{figure=fig1a.eps,height=60mm,width=60mm} \hspace{1 cm}
\psfig{figure=fig1b.eps,height=60mm,width=60mm} \caption{The ratio $R_\pi(q^2)=
F_\pi(q^2)/(1-q^2/m^2_\rho)$ vs $q^2$,
in the SL region. Experimental data from Ref.~\cite{Baldini}.}
\end{center}
\end{figure}

The momentum components of the vector meson vertex
$\Lambda_{n}(k,P_n)$ in Eq.~(\ref{cur5}) and the pion vertex
$\overline{\Lambda}_{\pi'}(k+P_{\pi},P_{\pi^{\prime}})$ evaluated
at $k^-=q^--(q-k)^-_{on}$ in Eq.~(\ref{amande}) are related to the
valence component of the respective front-form wave function. The
momentum dependence of the vertex for a hadron $h$ (a VM or a pion) is
\begin{eqnarray}
\Lambda_{h}(k,P_{h}) ]_{[k^- = q^--(q-k)^-_{on}]}=
\frac{C_h}{P^+_h}\left[M^2_h - M^2_0\right] \psi_{h}(k^+, {\bf
k}_{\perp}; P^+_{h}, {\bf P}_{h \perp}) \ , \label{wfn}
\end{eqnarray}
 where $\psi_{h}(k^+, {\bf
k}_{\perp}; P^+_{h}, {\bf P}_{h \perp})$ is an eigenstate of a
QCD-inspired mass operator \cite{Pauli2001}. The vertex $
\Lambda_{\pi}(k,-P_{\pi})$ evaluated at $k^-=q^--(q-k)^-_{on}$
represents a three-body system, composed by a radiated pion plus a
$q \bar{q}$ pair, as
 described in Ref.~\cite{Ji2001}. Such a vertex is given by a
pseudoscalar coupling multiplied by a constant, fixed through the
charge normalization.

The vector decay constant $f_{Vn}$ is calculated from the
covariant expression $\epsilon_{\lambda}^{\mu} \sqrt{2} f_{Vn}=
\langle0|\overline{q}\gamma^{\mu}q|\phi_{n,\lambda}\rangle$ for
the plus component of the em current to minimize the contribution
from possible zero-modes. The valence wave function, Eq.~(\ref{wfn}), is
introduced in the formula for $f_{V,n}$ after integrating over
$k^-$ in the momentum loop. Another ingredient of our model is
the probability of the VM valence component, roughly estimated by
$\sim 1/\sqrt{2n + 3/2}$~\cite{Pacheco2004}.

\section{Results and Conclusion}

The parameters in the form factor calculation are: i) the
oscillator strength ($\omega =1.39~GeV^2$) and the constituent
quark mass ($m = .265 ~GeV$), constrained by the meson
spectrum~\cite{Pauli2001}, and ii) a single vector meson width
(equal to $0.15~ GeV$) for the vector mesons with $M_n$ $\ge$
$2.150 ~GeV$, while the experimental  widths are used in the other
cases.

The  results of our model for the pion form factor
 show an overall agreement
both with  the TL and SL data~\cite{Pacheco2004},
from -10 $(GeV/c)^2$ to 10 $(GeV/c)^2$. In Fig. 1, a very detailed
comparison of our model in the SL region is presented.
In particular, the ratio between the pion form factor, $F_\pi(q^2)$
 and a  monopole, i.e. $ R_\pi(q^2)= F_\pi(q^2)/(1-\frac{q^2}{m_{\rho}^2})^{-1}$,
is adopted in order to avoid a log plot and therefore appreciate the virtue and
drawbacks of our approach. In Fig. 1, on the left,
  $R_\pi(q^2)$ is shown over the whole range of the experimental data
 ( as collected in Ref. \cite{Baldini}).
 The ratio tends asymptotically to a constant in
agreement with the expected behavior in QCD~(see
Ref.\cite{Brodsky98}); on the right,  $ R_\pi(q^2)$
is compared with the low momentum
transfer data, in order to investigate the charge radius.

In summary, the model reproduces satisfactory the experimental
data in the space-like region, while the agreement with the data
in the time-like region is only found at the qualitative level.
The next step is to improve the quality of the description in the
time-like region, introducing isospin mixing and possible coupling
of the photon to $^3D_1$ mesons.

We thank to FAPESP and CNPq for financial support.

\end{document}